\documentclass[nouppercase]{kaia}
\usepackage{xcolor}
\usepackage{graphicx}
\usepackage{amsmath}
\usepackage[utf8]{inputenc}
\usepackage{algorithm}
\usepackage[noend]{algpseudocode}
\usepackage{booktabs}
\usepackage{footnote}
\usepackage{multirow}
\makesavenoteenv{tabular}
\makesavenoteenv{table}
\usepackage{subcaption}
\usepackage{amssymb}
\usepackage[font=large]{caption}
\usepackage{float}

\title{\textbf{GDCA}: \textbf{G}AN-based single image super resolution with \textbf{D}ual discriminators and \textbf{C}hannel \textbf{A}ttention}

\author{Thanh Nguyen}{a}
\author{Hieu Hoang}{a}
\author{Chang D. Yoo}{a}
\affiliation{Korea Advances Institute of Science and Technology (KAIST)}{a}

\begin{document}

\maketitle
\begin{abstract}
Single Image Super Resolution (SISR) is a very active research field. This paper addresses SISR by using GAN-based approach with dual discriminators and incorporate with attention mechanism. The experimental results show that GDCA can generate sharper and high pleasing images compare to other conventional methods. 
\end{abstract}

\section{Introduction}
\begin{figure*}[htb]
\centering
\includegraphics[width=\textwidth]{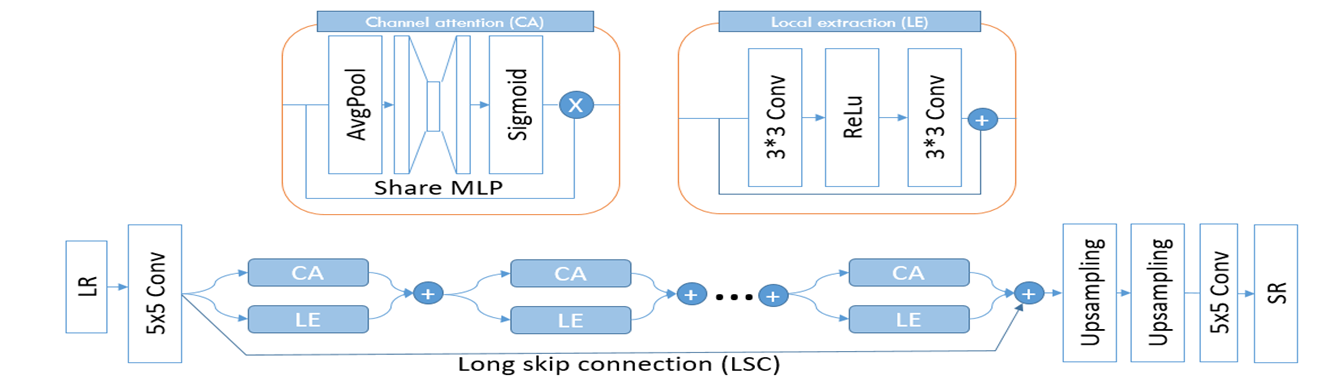}
\caption{The generator architecture consists of channel attention blocks and local extraction blocks with one long skip }\label{fig:arc2}
\end{figure*}

Single Image Super Resolution (SISR) is the problem of reconstructing an accurate high-resolution (HR) image from its low-resolution (LR) counterpart. The reconstructed image is referred as the super-resolution (SR) one. Recent approaches using deep learning show the impressive result with much higher quality compare to conventional approaches. This result engages more researchers to work on and make deep learning SISR to be an active research field along with other canonical topics \cite{nguyen2021pre,9413446,nguyen2021sample} . 

CNN based approach has achieved significant improvements over conventional methods. Dong et al. \cite{dong2014learning} is a pioneer with three-layer CNN (SRCNN) for the SISR. After that, VDSR \cite{kim2016accurate} was introduced with an increase of network depth to 20 layers, which outperforms SRCNN... These approaches aim to maximize Peak Signal Noise Ratio (PSNR) with the more efficient deeper network. The problem with these approaches is that the SR images are blurry result in visually unpleasing.

Recently, GAN-based methods have emerged as an effective solution to overcome the blurry limitation. Taking advantage of GANs enables to reconstruct SR images with high-frequency details and high perceptual quality. GAN based approach usually consists of a generator and a discriminator. Discriminator try to identify HR or SR image whereas generator try to fool discriminator to classify its generated SR image as HR image. SRGAN \cite{ledig2017photo} employs an adversarial loss term to increase visually pleasing quality. SRFeat \cite{park2018srfeat} used two discriminators and adopts the adversarial loss terms in both image and feature domains, resulting better perceptual quality.

Although previous approaches achieve high quality results, they did not take attention into account.. With SISR, attention can help the network focus on importance regions which have high texture in other to get higher performance.
In this work, we thoroughly incorporate attention with previous approaches and prove the effectiveness of attention mechanism in SISR. To be more specific, we propose “GAN-based single image super resolution with Dual discriminator and Channel Attention.” (GDCA).  GDCA takes advantage of GAN. GDCA introduce new generator with integrated attention module to boost performance. we also employ two discriminators which are image discriminator and feature discriminator. With these two discriminators, network discriminates in both image space and feature space resulting in a better perceptual quality.
Our contributions are listed below:

\begin{itemize}
    \item Propose a new generator architecture with a combination of skip connection, channel attention, batch norm removal, and use mean absolute error loss rather than conventional mean square loss.
    \item Dual discriminators are utilized: image discriminator and feature discriminator. The network is trained under GAN framework.
    \item We thoroughly evaluate our network on Root Mean Square Error (RMSE), “Perceptual Index” based on Ma + NIQE (PIMR2018). Experimental results show that our GDCA method is superior to has high performance on several benchmarks proving the effectiveness of attention in SISR. GDCA also generated much more pleasure image. The finding will engage people apply attention to their network in other to get high result on SISR.
\end{itemize}

\section{METHODOLOGY}
In this section, we present our proposed generator architecture as well as the explanation for our design. Then we explain the training loss functions used to train this network efficiently.

\begin{figure*}[htb]
\centering
\includegraphics[width=\textwidth]{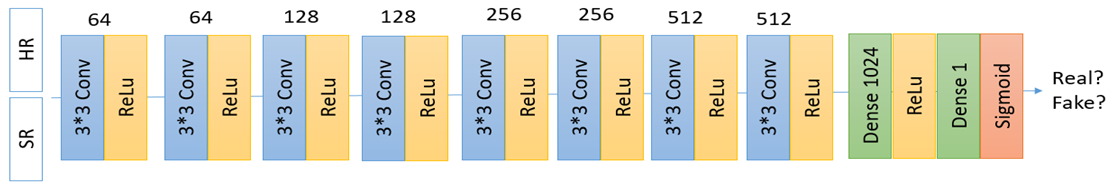}
\caption{The discriminator architecture. The number above a convolution represents the number of filters}\label{fig:arc1}
\end{figure*}

\subsection{Network architecture}

As the GAN-based architecture, our architecture includes two parts: generator and discriminators. Therein, the discriminators includes image discriminator and feature discrimination.
The generator network is shown in Figure \ref{fig:arc2}. The network receives the low resolution (LR) image and produces corresponding super resolution (SR) image. The overall network consists of multiple channel attention (CA) blocks and local extraction (LE) blocks with a weighted long skip connection. Firstly, a 5x5 convolution layer to extract low-level features are applied to input to extract course feature representation. Secondly, multiple CA blocks  and LE blocks are employed to learn higher-level features with more non-linearity and larger receptive fields. Lastly, two sub-pixel convolution layers (SPL) proposed in \cite{shi2016real} are used to up-sample the feature map to produce the target output.
The discriminators include feature discriminator and image discriminator. Both discriminators have the same network architecture but receive different inputs. The network architecture is combination of feed forward convolution neural network and fully connected layer as shown in Figure \ref{fig:arc2}. Image discriminator inputs are SR image and HR image whereas feature discriminator inputs are the feature map of SR image and HR image. These feature maps are extracted from corresponding Conv5 layer of VGG19 pre-train network of SR image and HR image. Both of the discriminators try to classify the image into real and fake class.
Utilizing the dual discriminator is the crucial factor of the whole framework. Image discriminator distinguishes the image space domain whereas feature discriminator distinguishes image in the feature domain. Both of discriminators make network stronger conventional discriminator result in a better output.

\subsection{Loss function}

There are three loss terms contributing to the total loss: perceptual similarity loss, image GAN loss, and feature GAN loss. Our losses follow the general GAN loss framework and aim to improve the perceptual quality.

\scalebox{0.8}{$\min _{g} \max _{d}\left(E_{y \sim p_{\text {data }}(y)}[\log (d(y))]+E_{x \sim p_{\text {data }}(x)}[\log (1-d(x))]\right)$}

Where g(x) is the generator network, d is a discriminator network, x is the input, and y is the sample from real data distribution.

\section{EXPERIMENTAL AND RESULT}
\subsection{Quantitative result}

We evaluated our GAN-based final generator on test set: Set 5. The final generator here was obtained by training the pre-train generator with GAN-based losses. Root Mean Square Error (RMSE) and Perceptual Index (PI) are chosen. These methods measure the perceptual quality of the output image. The higher PI, the better perceptual quality. Contrarily, the lower RMSE, the better reconstruction quality. The detail result is show in Table \ref{tab:res1}.

\begin{table}[t]
\centering
\small
\setlength{\tabcolsep}{4.5pt} 
\begin{tabular}{c|ccc}
\toprule
    & Super-SR\cite{dong2014learning}   & SRFeat \cite{park2018srfeat} & Ours \\ \midrule
Perceptual Index & 1.98 & 2.25 & \textbf{2.47} \\  \midrule
RMSE & 15.30  & 14.95  & \textbf{13.95}  \\\bottomrule
\end{tabular}
\caption{Result of our method compared to Super-SR and SRFeat on Set 5 evaluation set. Perceuptual Index and RMSE are measurement metrics.}
\label{tab:res1}
\end{table}

From the table, our generator produces better RMSE than other state of the art models, while obtains a com-parable perceptual index (2.47).

\subsection{Qualitative result}
\begin{figure}[htb]
\centering
	\begin{subfigure}{0.32\linewidth}
	\centering
		\includegraphics[width=\textwidth]{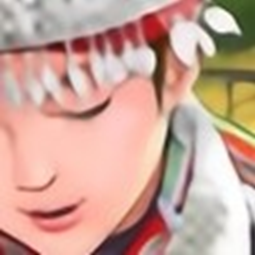}
		\caption{Mean Square}
	\end{subfigure}
	\begin{subfigure}{0.32\linewidth}
	\centering
		\includegraphics[width=\textwidth]{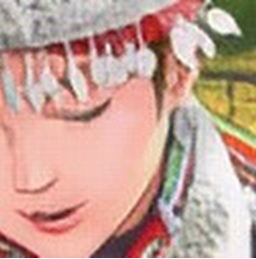}
		\caption{GDCA}
	\end{subfigure}
	\begin{subfigure}{0.32\linewidth}
	\centering
		\includegraphics[width=\textwidth]{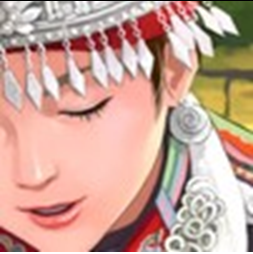}
		\caption{Ground truth}
	\end{subfigure}
\caption{Qualitative result performing on Set 5 test set. GDCA shows sharper result comparing to Mean Square-based method}\label{fig:res2}
\end{figure}

Our method show impressive qualitative result as show in Figure \ref{fig:res2}. Our GAN-based method achieves sharper output while mean square-based methods produce blurry result.

\section{CONCLUSION}
GDCE addresses SISR by using dual discriminators and incorporate with attention mechanism. The experimental results show that GDCA can generate sharper and high pleasing images compare to other conventional methods.

\bibliography{main}

\begin{thebibliography}{1}

\bibitem{dong2014learning}
Chao Dong, Chen~Change Loy, Kaiming He, and Xiaoou Tang.
\newblock Learning a deep convolutional network for image super-resolution.
\newblock In {\em European conference on computer vision}, pages 184--199.
  Springer, 2014.

\bibitem{kim2016accurate}
Jiwon Kim, Jung~Kwon Lee, and Kyoung~Mu Lee.
\newblock Accurate image super-resolution using very deep convolutional
  networks.
\newblock In {\em Proceedings of the IEEE conference on computer vision and
  pattern recognition}, pages 1646--1654, 2016.

\bibitem{ledig2017photo}
Christian Ledig, Lucas Theis, Ferenc Husz{\'a}r, Jose Caballero, Andrew
  Cunningham, Alejandro Acosta, Andrew Aitken, Alykhan Tejani, Johannes Totz,
  Zehan Wang, et~al.
\newblock Photo-realistic single image super-resolution using a generative
  adversarial network.
\newblock In {\em Proceedings of the IEEE conference on computer vision and
  pattern recognition}, pages 4681--4690, 2017.

\bibitem{9413446}
Thanh Nguyen, Tung Luu, Trung Pham, Sanzhar Rakhimkul, and Chang~D. Yoo.
\newblock Robust maml: Prioritization task buffer with adaptive learning
  process for model-agnostic meta-learning.
\newblock In {\em ICASSP 2021 - 2021 IEEE International Conference on
  Acoustics, Speech and Signal Processing (ICASSP)}, pages 3460--3464, 2021.

\bibitem{nguyen2021pre}
Thanh Nguyen, Tung~M Luu, Thang Vu, and Chang~D Yoo.
\newblock A pre-training framework for learning feature representation in
  visual observation reinforcement learning.
\newblock {\em Conference of IEIE}, pages 1890--1893, 2021.

\bibitem{nguyen2021sample}
Thanh Nguyen, Tung~M Luu, Thang Vu, and Chang~D Yoo.
\newblock Sample-efficient reinforcement learning representation learning with
  curiosity contrastive forward dynamics model.
\newblock {\em arXiv preprint arXiv:2103.08255}, 2021.

\bibitem{park2018srfeat}
Seong-Jin Park, Hyeongseok Son, Sunghyun Cho, Ki-Sang Hong, and Seungyong Lee.
\newblock Srfeat: Single image super-resolution with feature discrimination.
\newblock In {\em Proceedings of the European Conference on Computer Vision
  (ECCV)}, pages 439--455, 2018.

\bibitem{shi2016real}
Wenzhe Shi, Jose Caballero, Ferenc Husz{\'a}r, Johannes Totz, Andrew~P Aitken,
  Rob Bishop, Daniel Rueckert, and Zehan Wang.
\newblock Real-time single image and video super-resolution using an efficient
  sub-pixel convolutional neural network.
\newblock In {\em Proceedings of the IEEE conference on computer vision and
  pattern recognition}, pages 1874--1883, 2016.

\end{thebibliography}

\end{document}